\documentclass[aps,twocolumn,superscriptaddress,floatfix,prl]{revtex4-1}

\usepackage{lipsum}
\usepackage{graphicx}
\usepackage{amsmath,amssymb}
\usepackage{braket}
\usepackage{mathtools}
\usepackage{hyperref}
\usepackage{multirow}
\usepackage{color}
\usepackage{comment}
\usepackage[normalem]{ulem}
\usepackage{amsfonts}
\usepackage{float}
\usepackage{pdfpages} 
\makeatletter
\usepackage[caption=false]{subfig}

\def\beq{\begin{equation}}
\def\eeq{\end{equation}}
\def\bea{\begin{eqnarray}}
\def\eea{\end{eqnarray}}

\AtBeginDocument{\let\LS@rot\@undefined}
\makeatother

\begin{document}

\title{Thermalization of interacting quasi-one-dimensional systems}

\author{Mi\l{}osz Panfil}
\affiliation{Faculty of Physics, University of Warsaw, ul. Pasteura 5, 02-093 Warsaw, Poland}

\author{Sarang Gopalakrishnan}
\affiliation{Department of Physics, The Pennsylvania State University, University Park, PA 16802, USA}
\affiliation{Department of Electrical and Computer Engineering, Princeton University, Princeton, NJ 08544, USA}

\author{Robert M. Konik}
\affiliation{Condensed Matter Physics and Materials Science Division, Brookhaven National Laboratory, Upton, NY 11973, USA}

\begin{abstract}

Many experimentally relevant systems are quasi-one-dimensional, consisting of nearly decoupled chains.  In these systems, there is a natural separation of scales between the strong intra-chain interactions and the weak interchain coupling. When the intra-chain interactions are integrable, weak interchain couplings play a crucial part in thermalizing the system. Here, we develop a Boltzmann-equation formalism involving a collision integral that is asymptotically exact for \emph{any} interacting integrable system, and apply it to develop a quantitative theory of relaxation in coupled Bose gases in the experimentally relevant Newton's cradle setup. We find that relaxation involves a broad spectrum of timescales. We provide evidence that the Markov process governing relaxation at late times is \emph{gapless}; thus, the approach to equilibrium is generally non-exponential, even for spatially uniform perturbations.

\end{abstract}
\vspace{1cm}

\maketitle

The dynamics of thermalization---the approach to equilibrium of a quantum system initialized far from equilibrium---is a central theme in contemporary many-body physics~\cite{rigol_review}. This dynamics is particularly rich in one dimension, since many paradigmatic models, such as the Hubbard, Heisenberg, and Lieb-Liniger models, are integrable~\cite{takahashi2005thermodynamics}. Integrable models do not thermalize in the conventional sense, since they have extensively many local conserved densities; rather, they approach generalized Gibbs ensembles~\cite{rigol2008thermalization, gge_prl, langen2015experimental} that can have strikingly different properties (e.g., persistent charge and heat currents) from the standard Gibbs ensemble. Realistic experiments, especially in solid-state systems such as spin chains, never involve perfectly one-dimensional systems; the typical situation is that of a quasi-one-dimensional geometry of weakly coupled chains. One does not expect the system of coupled chains to be integrable; nevertheless, quasi-one-dimensional systems can feature a wide separation of scales between the intra-chain interactions---which generate the short-time dynamics---and the interchain interactions, which break integrability and thermalize the system. When this separation of scales is well-developed (as in cold atoms~\cite{PhysRevX.8.021030, kao2021topological}, as well as solid-state magnets~\cite{scheie2021detection}), the short-time dynamics is that of the integrable system, and at late times one sees a crossover to thermalization driven by the inter-chain couplings.

\begin{figure}[ht]
    \includegraphics[width = 0.45\textwidth]{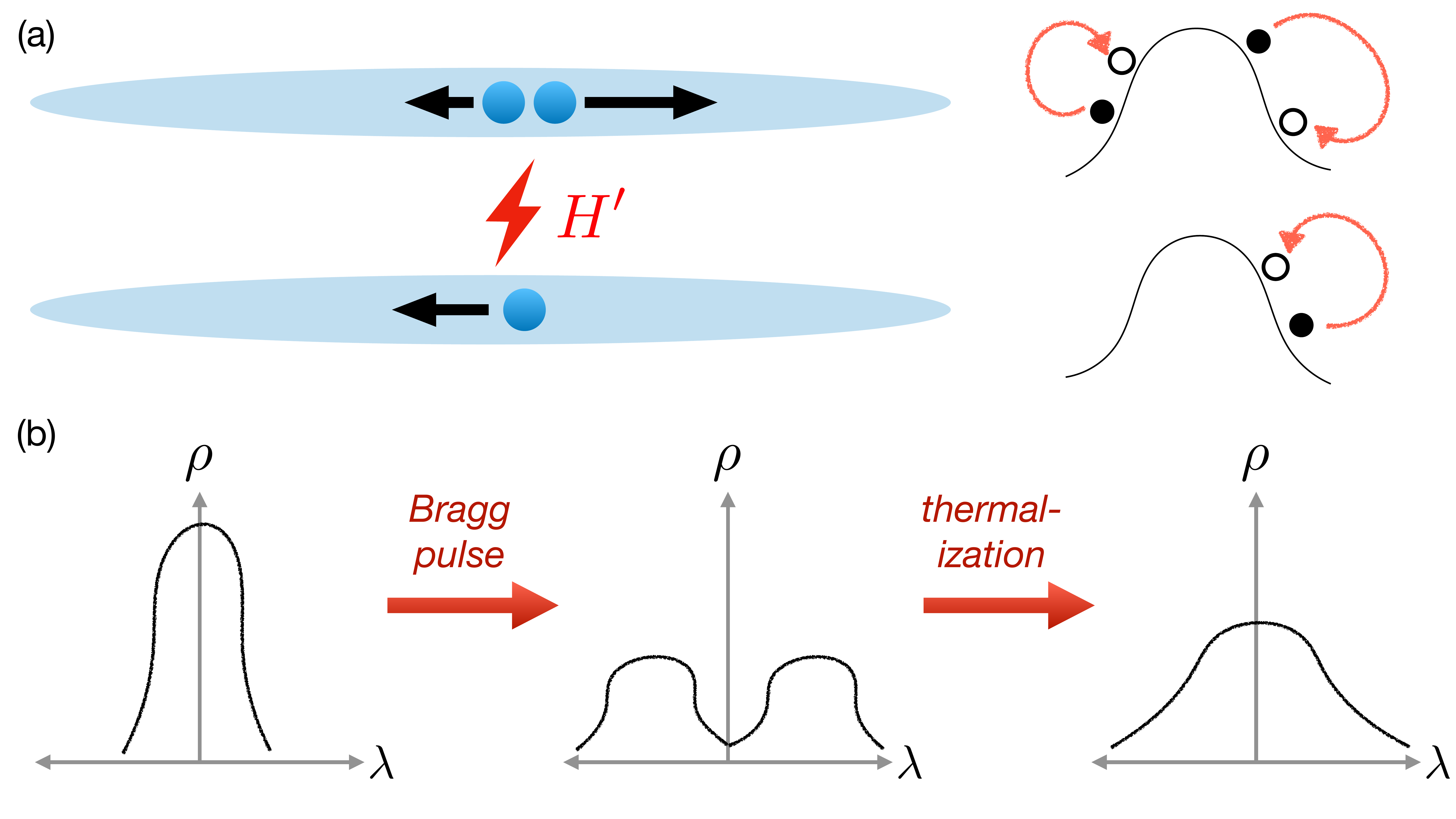}
    \caption{(a)~Depiction of two tubes coupled by the perturbation $H'$. The leading contribution to thermalization comes from collisions among three quasiparticles, which rearranges the rapidity distribution (sketched on the right side of the figure). (b)~Schematic for the evolution of the rapidity distribution under the experimental protocol, which consists of a Bragg pulse followed by thermalization.}
    \label{fig1}
\end{figure}

In the present work we address thermalization in such weakly coupled interacting integrable chains, initialized in arbitrary (but spatially uniform) nonequilibrium states. In this sense our work is complementary to the generalized hydrodynamics (GHD) program, which focuses on the relaxation of initially nonuniform states~\cite{doyon_ghd, fagotti_ghd, vir_ghd,10.21468/SciPostPhysCore.1.1.002, doyon2020lecture, PhysRevLett.122.090601, malvania2021generalized, PhysRevLett.126.090602}. We study the dynamics of thermalization via the time-evolution of the quasiparticle rapidity distribution, which is experimentally measurable through time-of-flight experiments~\cite{wilson2020observation}. The rapidity distribution evolves according to a Boltzmann equation, with a collision integral for which we develop an efficient, \emph{quantitatively} accurate computational scheme. Finding such explicit collision integrals has been one of the persistent challenges in the study of nearly integrable models: collision integrals that are similar to the one we derive were previously proposed in the literature~\cite{caux2019hydrodynamics, PhysRevB.101.180302, PhysRevB.102.161110, PhysRevLett.125.240604, weak_integ_breaking, bastianello2021hydrodynamics, PhysRevB.103.L060302, PhysRevLett.127.057201}, but have not been used to study relaxation from physically relevant nonequilibrium initial states. (For complementary approaches to the problem of weak integrability breaking see Refs.~\cite{PhysRevLett.115.180601, PhysRevB.94.024506, PhysRevX.9.021027, bulchandani, PhysRevX.10.041017,2021PhRvE.103d2121H, PhysRevB.104.L201117}.) Here, we apply our collision integral to characterize the relaxation of coupled Bose gases initialized in the experimentally relevant ``Newton's cradle'' setup~\cite{kinoshita2006quantum} (Fig.~\ref{fig1}). In an (idealized) Newton's cradle experiment, a gas is prepared in a low-temperature equilibrium state, and is then subjected to a pulse that boosts the momenta of half the atoms by $p$ and the other half by $-p$. When the dynamics is not exactly integrable~\cite{PhysRevX.8.021030}, this nonequilibrium state slowly relaxes to a higher-temperature equilibrium state (Fig.~\ref{fig1}b).

Our main result is a quantitative description of how the quasiparticle distribution evolves during this relaxation process. From the quasiparticle distribution, we can also straightforwardly compute the evolution of charge and energy currents, and of the entire hierarchy of charges that are strictly conserved in the integrable limit~\cite{gge_prl, PhysRevLett.119.020602}. For the far-from-equilibrium initial state we focus on, this relaxation is a complex multiple-scale process that we can only solve numerically. However, assuming the system thermalizes, then at \emph{late} times it is near equilibrium and one can linearize the Boltzmann equation. We present evidence that the spectrum of the resulting linear operator is gapless and spans several orders of magnitude.  The late-time approach to equilibrium is thus not governed by a single characteristic timescale. This is true, remarkably, \emph{even though} the initial state is spatially uniform, so the quench does not directly couple to any hydrodynamic-scale density fluctuations.

\emph{Model}---We consider an array of one-dimensional bosonic gases (``tubes''), oriented along the $x$ axis, each governed by the Lieb-Liniger Hamiltonian
\beq
H_{\mathrm{LL}, i} = \int dx\, \hat{\psi}_i^\dagger(x) \left[ - \frac{1}{2m} \partial_x^2 + c \hat{\rho}_i(x) \right] \hat{\psi}_i(x).
\eeq
Here, $i$ indexes the tubes, $m$ is the microscopic mass of the bosons, $\hat{\rho}_i(x) \equiv \hat{\psi}^\dagger_i(x) \hat{\psi}_i(x)$ is the density operator, and $c$ is a coupling constant.
The tubes are coupled to one another by density-density interactions of the form
\beq
H' = V_0 \sum_{ij} \int dx dx' A_{ij}(|x - x'|) \hat{\rho}_i(x) \hat{\rho}_j(x').
\eeq
We will leave $A_{ij}(|x - x'|)$ generic for now, but we are primarily interested in the case of dipole-dipole interactions, where $A_{ij} = (1 - 3 \cos^2(\theta_{ij}))/(|x - x'|^2 + |\mathbf{r}_i - \mathbf{r}_j|^2)^{3/2}$. Here, $\mathbf{r}_i \equiv (y_i, z_i)$ is the position of the $i$th tube in the array, and $\theta_{ij}$ is the angle between the separation $\mathbf{r}_i - \mathbf{r}_j$ and the orientation of the dipoles (which is fixed in the experiment by applying a magnetic field). 

For simplicity we anticipate that $A_{ij}$ falls off fast enough with distance between tubes that it is sufficient to consider nearest-neighbor interactions between tubes. Thus each tube interacts with $z$ neighbors. With the Newton's cradle experiment in mind, we also assume in what follows that the tubes have identical quasiparticle distributions in the nonequilibrium initial state.
Under these assumptions, it suffices to consider the effects of the integrability-breaking perturbation acting on a pair of neighboring tubes; we thus drop the indices on the interaction shape~$A(x)$.

\emph{Boltzmann equation for two tubes}---We assume a separation of scales between the fast dynamics due to $H_{\mathrm{LL}}$ and slow dynamics due to $H'$.  Without loss of generality we pick tube 1 as the ``system'' tube (whose rapidity distribution is being measured) and tube 2 as a ``bath'' tube. On timescales that are long compared with the fast dynamics, the state of each tube can be characterized by a generalized Gibbs ensemble~\cite{gge_prl,2012PhRvL.109q5301C,2012_Mossel_JPA_45}, or equivalently by its quasiparticle distribution function $\rho_p(\lambda)$, where $\lambda$ is the ``rapidity''. The rapidity labels particles in interacting integrable models in an analogous way to momentum in free theories. The distribution $\rho_p(\lambda)$ evolves in general~as
\beq
\partial_t \rho_{p,1}(\lambda) = \tau^{-1} Q[\rho_{p,1},\rho_{p,2}] (\lambda). \label{master_eq}
\eeq
We emphasize that the right-hand side (the so-called ``collision integral'') is a nontrivial functional of the density distributions in the two tubes. For the present, we restrict to the case where the density distributions are initially identical, and thus stay identical at all times (at our level of analysis). The time scale for the evolution follows from the Fermi's golden rule and is set by $\tau^{-1} \equiv 16 E_F/(\hbar \pi^2)  \times \gamma_{\rm inter}^2$, where $E_F$ is the Fermi energy of the single tube and $\gamma_{\rm inter} \equiv V_0 m/(n_{\rm 1D}\hbar^2)$ is the dimensionless coupling between the two tubes with $n_{\rm 1D}$ the one-dimensional density of the gas. The full derivation of the collision integral, including the case of different distributions in the two tubes, is presented in the Supplemental Material, Section $2$. Here we discuss in details the ingredients of the resulting expression.

The population at rapidity $\lambda$ may change for two reasons: either because a particle directly scatters into or out of that rapidity, or indirectly due to interactions. (As an example, in a finite system, changing the rapidity of one particle alters the quantization condition for all the others, via the Bethe equations.) In the thermodynamic limit, this ``backflow'' effect can be taken into account through the relation $Q[\rho_p] (\lambda) = \int d\mu R[\rho_p] (\lambda, \mu) Q_0[\rho_p] (\mu)$, where $Q_0$ is the direct scattering rate given by Fermi's Golden Rule, and $R$ is an integral operator (which is purely a property of the integrable dynamics) that captures the influence of this scattering process elsewhere in rapidity-space, see \footnote{Please see Section S2 of the Supplementary Material.}. Henceforth we will drop the $\rho_p$ argument, noting that all quantities of interest are functionals of the full distribution. A physical choice of $Q_0$ must conserve particle number in each tube, as well as total momentum and energy in the full array. In our setup, since the tubes are identical, momentum and energy will be also conserved on `average' (in a temporal sense) in each tube.

We now turn to computing $Q_0$. 
We write $Q_0 = \sum_{n, m} Q_0^{(n,m)}$, where $Q_0^{(n,m)}$ is a scattering process involving $n$ particle-hole excitations in the system tube and $m$ particle-hole excitations in the bath tube. 
We can write $Q_0^{n,m}(\lambda)$ so that the focus is on the n-particle-hole excitations in the system tube.  Let these be indexed by $\{(p_i,h_i)\}^n_{i=1}$.  We want to consider all possible distinct particle-hole combinations such that the rapidity $\lambda$ is equal to $p_i$ or $h_i$ for some i.  The scattering event, $\{(p_i,h_i)\}^n_{i=1}$, will change the system tube's energy/momentum by $(\omega,k)$.  By energy-momentum conservation the $m-$particle-hole process in the bath tube must change by $(-\omega,-k)$.  The ability of the bath tube to support such an excitation complex is encoded in the spectral density of the bath tube, a quantity we denote as $S^m_2(k,\omega)$.  The scattering processes in both tubes are governed by matrix elements of the density operator (as the interaction between tubes is density-density).  We write the matrix element for the system tube as $F^{\rho_1}(\{ p_i, h_i \}^m_{i=1})= \langle \rho_p | \hat \rho_1(x) | \rho_p; \{ h_i \} \to \{ p_i \} \rangle$.  The contribution of the particle-hole combination $\{(p_i,h_i)\}^n_{i=1}$ in the system tube to $Q_0^{n,m}$ is controlled by the particle and hole densities, $\rho_{p1}$/$\rho_{h1}$ and so is proportional to $\prod_{i=1}^{n}\rho_{p_1}(h_i)\rho_{h_1}(p_i)$.  Putting these features together allows us to write $Q_0^{n,m}$ as 
\begin{widetext}
\begin{eqnarray}\label{gen_boltzmann}
Q^{(n,m)}_0(\lambda) &=& \frac{n}{(n!)^2} \int \prod_{i = 1}^{n} d p_i d h_i \delta(\lambda - p_1) A^2(k) \left|F^{\rho_1}(\{ p_i, h_i \})\right|^2 \left[ \rho_{p1}(h_i) \rho_{h1}(p_i) S_2^{m}(-k, -\omega)- ({h_i}\leftrightarrow{p_i}) \right].
\end{eqnarray}
\end{widetext}
We derive this form more concretely in \footnote{Please see Section S2.2-S2.3 of the Supplementary Material.}.

To make further progress with Eq.~\eqref{gen_boltzmann} we must evaluate the $m$-particle-hole matrix elements $F^{\rho}$ and the $m$-particle-hole contribution to the dynamic structure factor. 
In general, processes with any $(n,m)$ contribute comparably to relaxation and one must sum over these processes, which is evidently intractable. However under the assumption that the intertube interactions are varying smoothly with the distance the relaxation is dominated by processes transferring small momenta. This implies that higher particle-hole processes that involve higher powers of the momentum transfer (see Appendix A), can be neglected. Another regime dominated by few-particle processes is the $c \to \infty$ (Tonks-Girardeau) limit in which processes involving $n-$particle-holes are suppressed by~$c^{-2n}$.

The simplest interaction process is governed by $Q_0^{(1,1)}$, i.e., by two-particle scattering. However, the kinematics of one-dimensional two-body scattering is too restrictive to lead to thermalization; indeed, if the distributions in the two tubes are initially the same, $Q_0^{(1,1)}$ has no nontrivial dynamical effects. %
Thus, the leading processes that do contribute are $Q_0^{(1,2)}$ and $Q_0^{(2,1)}$: i.e., diffractive three-body scattering processes involving two particles in one tube and one in the other~\footnote{These two processes are physically different: in $Q^{(1,2)}$, the quasiparticle with rapidity $\lambda$ scatters to a new state while causing a two-particle rearrangement in the other tube; in $Q^{(2,1)}$ this quasiparticle is involved in a two-body scattering process.}.

\begin{figure}[b!]
	\includegraphics[width=\linewidth, trim=7cm 5cm 4cm 4cm, clip]{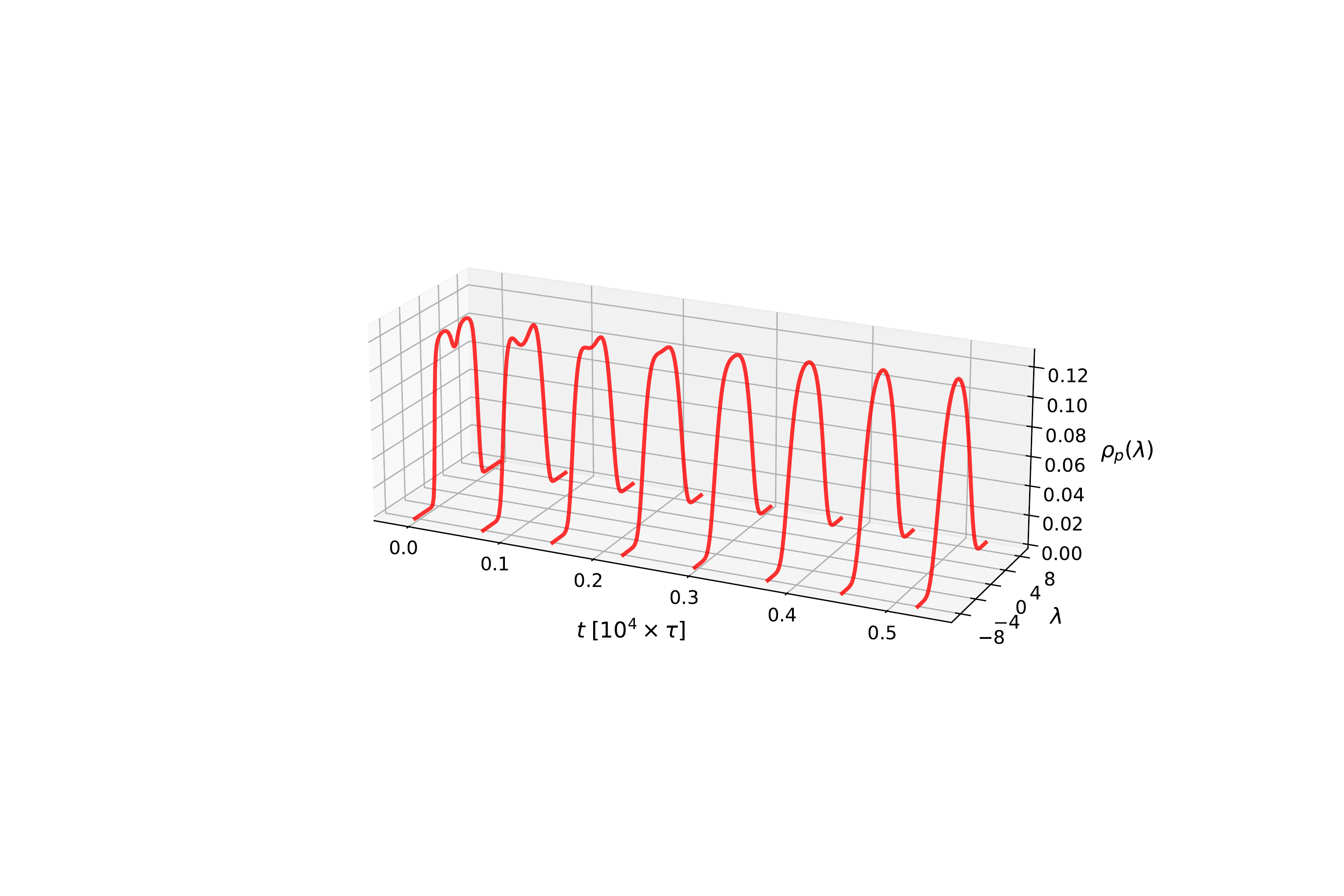}
	\caption{Evolution of the quasiparticle distribution in one of the tubes for the nonequilibrium protocol described in the main text. The evolution is characterized by a quick washing out of the two Bragg peaks followed by a relatively slower approach to the final equilibrium distribution, as further shown in Fig.~\ref{fig:th}(a).} \label{fig:den_ev}
\end{figure}

The scattering rates $Q_0^{(1,2)}$ and $Q_0^{(2,1)}$ can be evaluated in the limit of small momentum transfer, using recently developed expressions for the form factors $F^\rho(p_; h)$ and $F^\rho(p_1, p_2; h_1, h_2)$ above a generalized Gibbs state~\cite{Smooth_us,SciPostPhys.1.2.015,2018JSMTE..03.3102D,10.21468/SciPostPhys.6.4.049,Bootstrap_JHEP,Cortes_Cubero_2020,milosz_2021}. The dynamic structure factor can also be expressed in terms of these same form factors by means of a spectral representation \cite{esskon}.  One of the challenges in employing form factors in computing the scattering rates and structure factors is to make sense of the non-integrable singularities they introduce \cite{kon2003,esskon2009,PhysRevB.78.100403}.  
In order to tackle this problem we use the Hadamard regularization~\footnote{Please see Section S5 of the Supplementary Material.}, the method used earlier in computing response functions at finite energy density as well as in the context of diffusion in generalized hydrodynamics, Refs.~\onlinecite{Pozsgay_2010,DeNardis2019,Bootstrap_JHEP,milosz_2021}.

\emph{Results}---

\begin{widetext}

\begin{figure}[t]
    \subfloat{\includegraphics[width=0.34\textwidth]{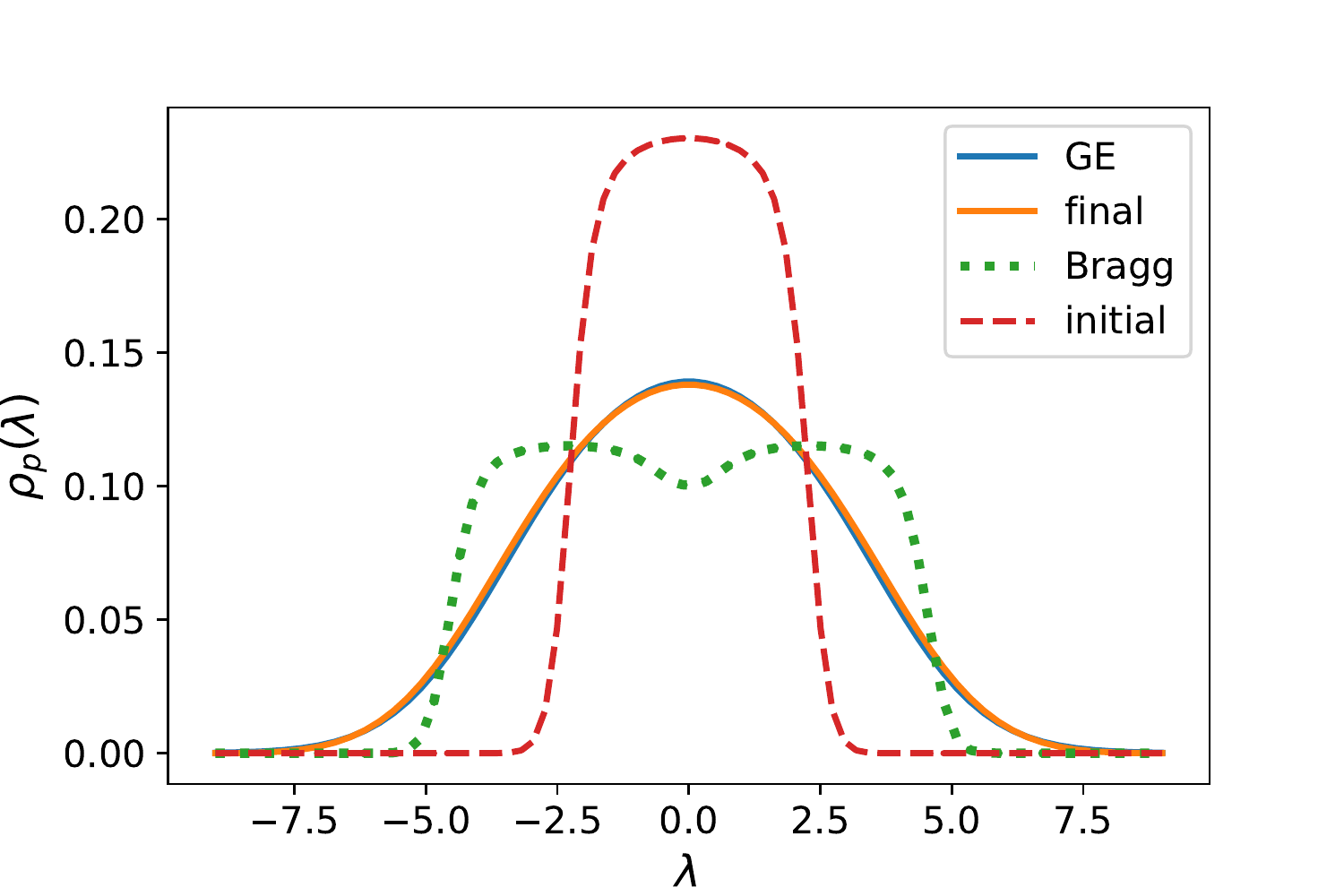}}
	\subfloat{\includegraphics[width=0.34\textwidth]{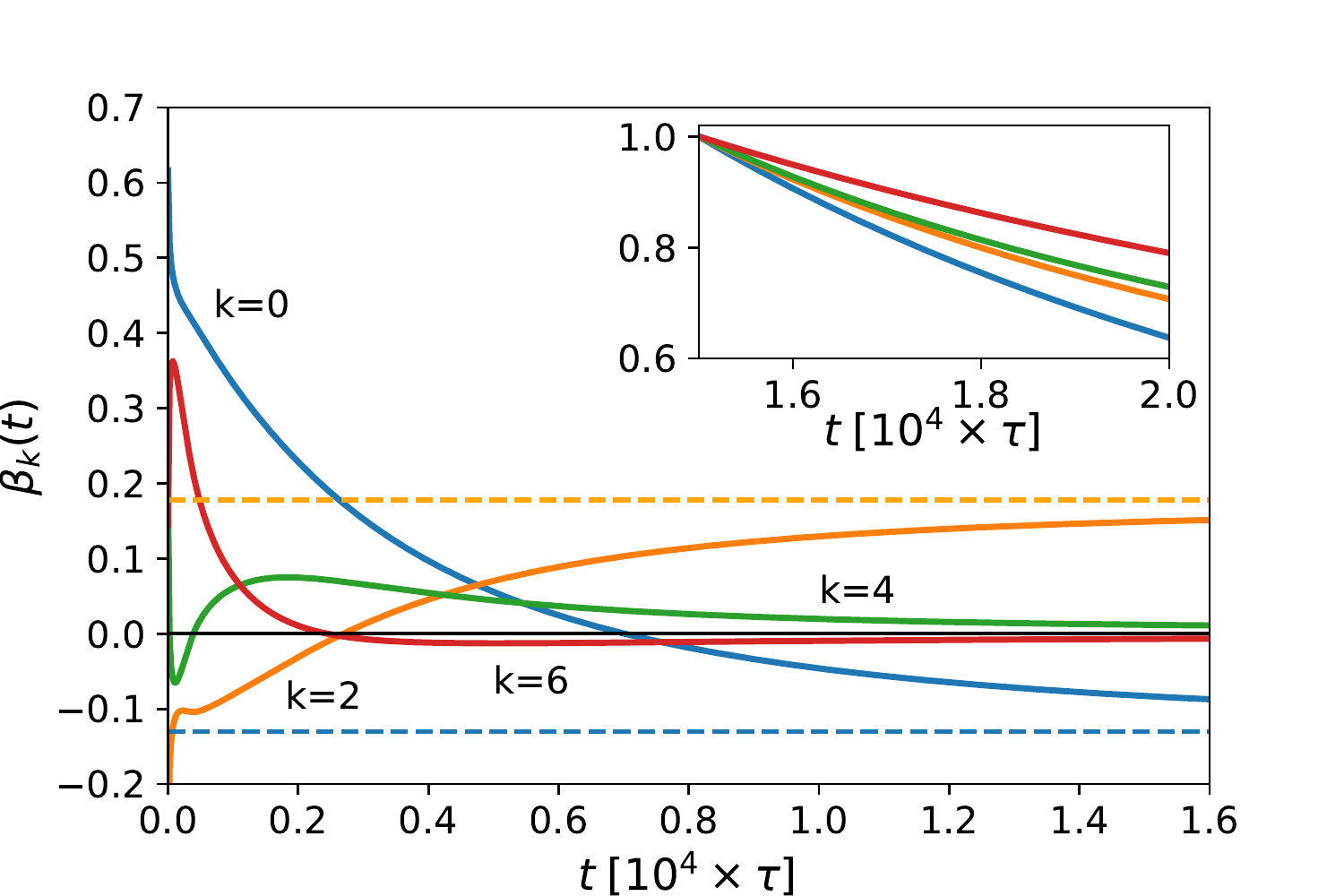}}
    \subfloat{\includegraphics[width=0.34\textwidth]{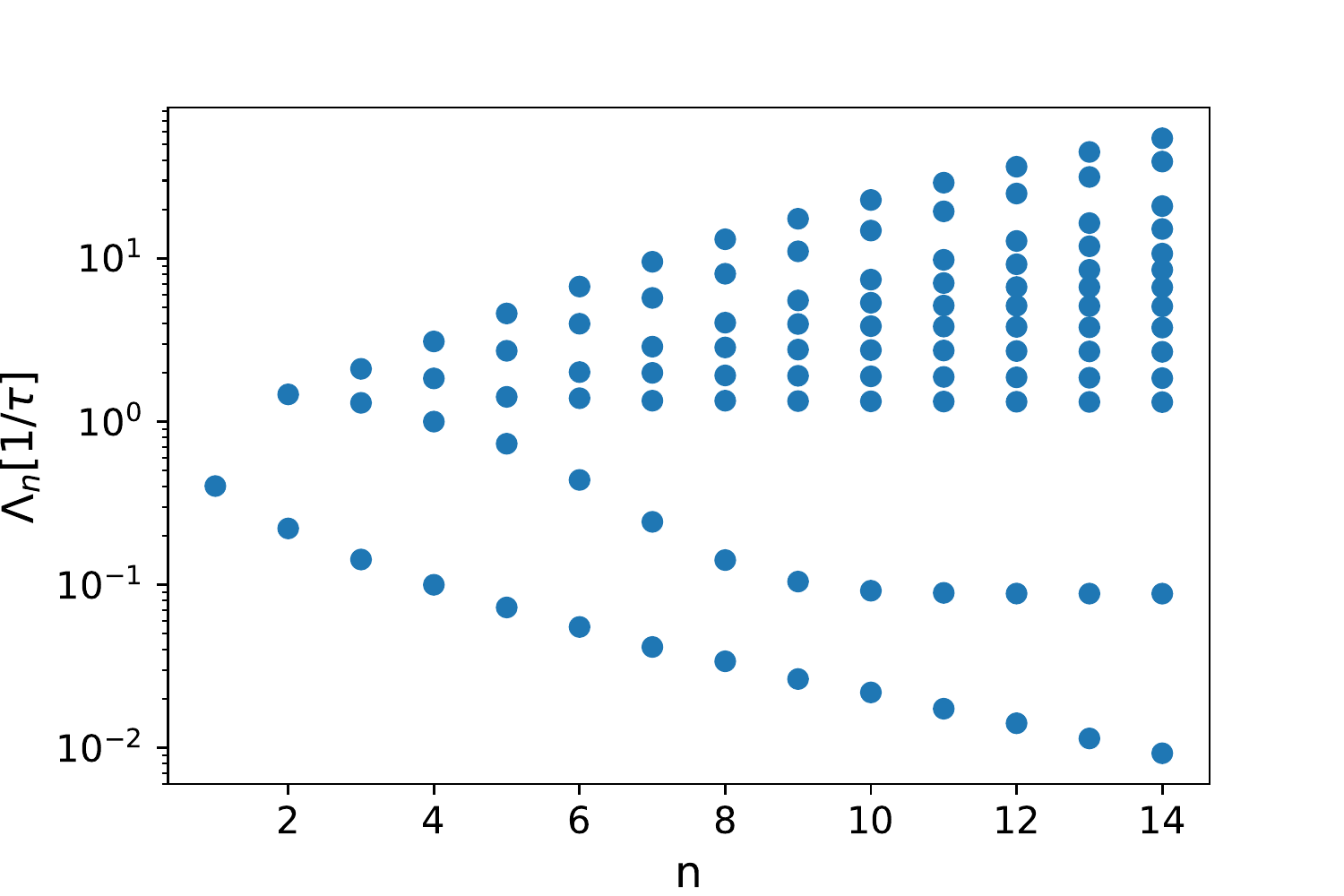}}
    \caption{(a) Initial, post Bragg-pulse and final (at $t = 2 \times 10^4 \tau$) quasiparticle distributions. The final distribution coincides with the thermal equilibrium distribution with $k_BT \approx 11.25$ fixed from the energy of the post Bragg-pulse state.  (b) Evolution of the generalized chemical potentials $\beta_k(t)$ determining the ensemble of the gas as a function of time. We consider truncated GGE of ultra-local charges with first $4$ even charges. The dashed lines are the equilibrium values. In the inset we show that the thermalization rates for different chemical potentials are different. (c) The eigenvalues of the dimensionless linearized and truncated evolution operator $\bar{\mathbf{Q}}_0^{(1,2)}$ for the thermal state approached by nonequilibrium evolution and for different values of the truncation order $n$. To simplify the interpretation, while evaluating $\bar{\mathbf{Q}}_0^{(1,2)}$, we work in the large $c$ limit, where the dressings are subleading. All the eigenvalues are positive and the smallest and largest eigenvalues differ by few orders of magnitude. The smallest value monotonically decrease upon increasing the truncation order $n$ and conjecturally approaches zero.} \label{fig:th}
\end{figure}

\end{widetext}

We study now in details the time evolution of the system prepared in the following initial state motivated by the recent experiments~\cite{PhysRevX.8.021030}.  Initially, the two tubes are in thermal equilibrium at temperature $T_0$ and with the chemical potential $h_0$. The corresponding quasiparticle distribution is  $\rho_{T_0, h_0}(\lambda)$. We imagine now performing a Bragg pulse effectively boosting each cloud of atoms by $\pm p$ such that $\rho_{pi}(\lambda, 0) = (\rho_{T_0, h_0}(\lambda + p) + \rho_{T_0, h_0}(\lambda - p))/2$. The system then evolves according to~\eqref{master_eq}. As discussed above, for the distributions in both tubes are identical the leading processes are $(1,2)$ and $(2,1)$. To be specific we fix the interaction parameter $c=4$ which corresponds to a strongly correlated regime of the Lieb-Liniger model. For the initial state we choose system at $k_B T=1$ and unit density $n_{\rm 1D} = 1$ and set $p = 2.3$. In Fig.~\ref{fig:den_ev} we show the resulting time evolution and in Fig.~\ref{fig:th}(a) we compare the initial distribution and the distribution at late time. This clearly shows the thermalization with the thermal distribution fixed by the particle density and the energy right after the Bragg pulse. Additionally the thermalization process can be witnessed by observing the diagonal entropy production~\cite{2011AnPhy.326..486P} as we discuss in more details in the Supplementary Material, Section S3).

The instantaneous states of the system can be described by the generalized Gibbs ensemble (GGE). The GGE involves, beside particle number and total energy, also all other local conserved charges $Q_k$ present in uncoupled Lieb-Liniger model. The GGE density matrix takes then the form $\hat{\rho} \sim \exp(- \sum_{k} \beta_k Q_k)$. The distribution $\rho_{\rm p}(\lambda)$ is in one-to-one correspondence with the chemical potentials $\beta_k$ (see Appendix B). The dynamics of $\rho_{\rm p}(\lambda)$ can be then translated in the time dependence of the chemical potentials $\beta_k(t)$. This is a convenient way to explore thermalization: in the thermal state only $\beta_0$ and $\beta_2$ are not zero, so all other $\beta_n(t)$ should decay to zero.
The initial distribution is an even function of the rapidity and the time evolution does not modify that and therefore only even chemical potentials are potentially non-zero. In Fig.~\ref{fig:th}(b) we plot first few chemical potentials. We observe that $\beta_k(t)$ for $k > 2$ approach at large times zero signalling again the thermalization. In the inset we consider $(\beta_k(t) - \beta_k^{\rm th})$ for large times and normalized by the values at specific time $t^*$ such that each line starts at $1$. This shows that the system does not display a single timescale for thermalization. Instead the thermalization rate depends on the generalized chemical potential considered with the one corresponding to the particle number and energy evolving the slowest. 

At very late times, we can assume the system is near a thermal state, allowing us to linearize the Boltzmann equation about the thermal state. The process of thermalization is then determined by the spectrum of the resulting linear operator $\mathbf{Q}_0$~\footnote{Please see Section S4 of the Supplementary Material}. This operator further decomposes into contributions $\mathbf{Q}_0^{(n,m)}$ analogously to the full dynamics. The two leading processes $(1,2)$ and $(2,1)$ lead then to a symmetric operator $\bar{\mathbf{Q}}_0^{(1,2)} = \mathbf{Q}_0^{(1,2)} + \mathbf{Q}_0^{(2,1)}$ whose spectrum we now analyse.  

The spectrum of this operator has three zero modes, corresponding to energy, particle number, and momentum conservation.
The rate of approach to the steady state is set by the smallest-magnitude nonzero eigenvalue. 
To learn about its spectrum we truncate the infinite-dimensional operator $\bar{\mathbf{Q}}_0^{(1,2)}$ to a finite-dimensional space spanned by the lowest ultra local conserved charges not contained in its kernel. 
The spectrum of this truncated operator is plotted in Fig.~\ref{fig:th}(c) as a function of the truncation order. We find that the magnitude of the lowest nonzero eigenvalue rapidly decreases with increasing truncation order. Our numerical results suggest (though we cannot prove) that the operator is gapless: thus, there is a spectrum of relaxation times going all the way out to infinity, and the approach to the steady state is non-exponential. This feature is unexpected: usually, power-law relaxation in nonintegrable systems is associated with the hydrodynamics of long-wavelength density fluctuations, but the initial states we consider are translation-invariant and do not have such fluctuations. Understanding the origin of this gapless spectrum---and whether it is generic---is an interesting question for future work.

\emph{Summary}---In this work we addressed a central challenge in the study of nearly integrable systems: we developed a Boltzmann equation with a \emph{microscopically derived} collision integral, to describe the relaxation of the system to equilibrium. This Boltzmann equation is quantitatively accurate for perturbations that fall off slowly in space, e.g., dipolar interactions between integrable chains. This Boltzmann equation applies to arbitrary initial states, though for simplicity we assumed translation invariance. Our main result is that relaxation from the Newton's cradle setup is a process that involves many different timescales: indeed, our numerical results on the linearized Boltzmann equation suggest that relaxation is non-exponential even at the latest times.  An important future direction would be to extend our results to spatially inhomogeneous initial states and systems confined in harmonic traps: this would involve combining our Boltzmann equation (applied locally) with the nontrivial evolution of the quasiparticle distributions under generalized hydrodynamics, including space-time inhomogeneities~\cite{PhysRevLett.123.130602}. In experimental setups the gas is confined in a 3D trap. This leads to a new integrability breaking perturbation through virtual excitations into higher radial modes~\cite{PhysRevB.94.024506}. In the case considered here where the dynamics are dominated by low momenta particle-hole excitations, virtual excitations involving higher radial modes are suppressed by a high power of momentum ~\cite{PhysRevB.94.024506} and so presumably are subdominant.  Nonetheless, it would be valuable to understand the effects of such integrability breaking on dynamics in more general scenarios. We are also interested in applying the formalism developed herein to the problem of thermalization in spin chain materials \cite{scheie2021detection} including rare earth variants \cite{Wu2016,Gannon2019,PhysRevB.101.245150} placed out of equilibrium and probed by neutron and resonant inelastic x-ray scattering.

\emph{Acknowledgments}--- M.P. acknowledges the support from
the National Science Centre, Poland, under the SONATA
grant 2018/31/D/ST3/03588.
S.G. acknowledges support from NSF DMR-1653271.
R.M.K. was supported by the U.S. Department of Energy, Office of Basic Energy Sciences, under Contract No. DE-SC0012704.

\emph{Appendix A}---

In this appendix we analyze the dependence of the scattering integral on the momentum $k$ transferred between the tubes. 
To this end, the integration over $(p_i, h_i)$ in~\eqref{gen_boltzmann} can be transformed into an integration over $(k_i, \omega_i)$ where
\begin{equation}
    k_i = k(p_i) - k(h_i), \qquad \omega_i = \omega(p_i) - \omega(h_i).
\end{equation}
The Jacobian of the $J$ transformation is 
\begin{equation}
    J^{-1} = \prod_{j=1}^n k'(p_i) k'(h_i) | v^{\rm eff}(p_i) - v^{\rm eff}(h_i)|.
\end{equation}
We assume that in the small momentum limit the relevant excitations take the form of small particle-hole excitations, $p_i \sim h_i$. 
For small particle-hole excitations, we have that $J \sim \prod_{i=1}^n |k_i|^{-1}$. Therefore each integration over $k_i$ and $\omega_i$, including the presence of the Jacobian, gives a factor $k_i$ (we assume the energy is linear in $k_i$). The Dirac $\delta$-function reduces the number of integrals by one and effectively decreases the order in $k$ by $1$. Therefore, the phase space of the excitations scales like $k^{n-1}$. 

We use now that the form-factors are, in the leading order, momentum independent and that for $\omega \sim k$, $S^m(k, \omega) \sim |k|^{m-2}$~\cite{milosz_2021}. Finally, because $\rho_{p1}(h_i) \rho_{h1}(p_i) - ({h_i}\leftrightarrow{p_i}) \sim k$, there is an additional power of $k$ coming from the particle-hole distributions. For symmetric potential $A(k) = A(-k)$ the leading term vanishes and collecting the factors we find~$Q_0^{(n,m)} \sim k^{n+m-1}$.

\emph{Appendix B}--- 

To read off the chemical potentials $\beta_k$ from a given distribution $\rho_{\rm p}(\lambda)$ we invert the usual procedure of computing the distribution from the knowledge of the chemical potentials~\cite{2012PhRvL.109q5301C}. In practice we first compute $\rho_{\rm tot}(\lambda)$ from the defining integral equation~\cite{KBI},
\begin{equation}
 \rho_{\rm tot}(\lambda) = \frac{1}{2\pi} + \!\int\!{\rm d}\mu\, T(\lambda - \mu) \rho_{\rm p}(\mu), \; T(\lambda) = \frac{c}{\pi}\frac{1}{\lambda^2 + c^2}.
\end{equation}
This gives us an access to the filling function $n(\lambda) = \rho_{\rm p}(\lambda)/\rho_{\rm tot}(\lambda)$. The filling function is expressed through the pseudo-energy $\epsilon(\lambda)$ as $n(\lambda) = [1 + \exp(\epsilon(\lambda))]^{-1}$. The pseudo-energy itself is related to the bare pseudo-energy $\epsilon_0(\lambda)$ through the integral relation,
\begin{equation}
	\epsilon(\lambda) = \epsilon_0(\lambda) - \int {\rm d}\mu\, T(\lambda - \mu) \log \left(1 + e^{-\epsilon(\mu)} \right).
\end{equation}
Finally, the bare pseudo-energy is expressed through the chemical potentials as $\epsilon_0(\lambda) = \sum_k \beta_k \lambda^k$, where $\lambda^k$ is the single particle contribution to the charge $Q_k$, namely
\begin{equation}
	Q_k = \int {\rm d}\lambda\; \lambda^k \rho_{\rm p}(\lambda).
\end{equation}

\bibliography{bib}

\end{document}


\title{Supplemental Material: Thermalization of interacting Quasi-one-dimensional Systems}

\author{Mi\l{}osz Panfil}
\affiliation{Faculty of Physics, University of Warsaw, ul. Pasteura 5, 02-093 Warsaw, Poland}
\author{Sarang Gopalakrishnan}
\affiliation{Department of Physics, The Pennsylvania State University, University Park, PA 16802, USA}
\affiliation{Department of Physics and Astronomy, CUNY College of Staten Island, Staten Island, NY 10314, USA}
\author{Robert M. Konik}
\affiliation{Condensed Matter Physics and Materials Science Division, Brookhaven National Laboratory, Upton, NY 11973, USA}
\date{
    \today
}
\title{Supplemental Material: Thermalization of interacting Quasi-one-dimensional Systems}

\maketitle

\appendix

\setcounter{equation}{0}
\setcounter{figure}{0}
\setcounter{table}{0}
\makeatletter

\renewcommand{\theequation}{\thesection.\arabic{equation}}
\renewcommand{\thefigure}{S\arabic{figure}}

\renewcommand{\appendixname}{}

\renewcommand{\thesection}{S\arabic{section}}
\titleformat{\section}[hang]{\large\bfseries}{\thesection}{0.5em}{}
\titleformat{\subsection}[hang]{\bfseries}{\thesection.\thesubsection}{0.5em}{}
\def\p@subsection     {\thesection.}

\noindent Here we report some additional technical details on our work.

\section{Thermodynamics of the Lieb-Liniger model}

In this section we recall the necessary ingredients for describing thermodynamic states of the Lieb-Liniger model. We follow Ref.~\onlinecite{KBI}.

The central function characterizing the state of the gas is the filling function $n(\lambda)$. This function defines the quasiparticles distribution $\rho_{\rm p}(\lambda)$ through $\rho_{\rm p}(\lambda) = n(\lambda) \rho_{\rm tot}(\lambda)$ where
\begin{equation}
    \rho_{\rm tot}(\lambda) = \frac{1}{2\pi} + \int {\rm d}\lambda'\, T(\lambda, \lambda') n(\lambda') \rho_{\rm tot}(\lambda'),
\end{equation}
and the kernel $T(\lambda,\lambda')$ is defined to be
\begin{equation}
    T(\lambda,\lambda') \equiv \frac{1}{2\pi}K(\lambda-\lambda') = \frac{c}{\pi}\frac{1}{c^2+(\lambda-\lambda')^2}.
\end{equation}
The distribution of holes $\rho_{\rm h}(\lambda)$ is defined with respect to the total density, namely $\rho_{\rm h}(\lambda) = \rho_{\rm tot}(\lambda) - \rho_{\rm p}(\lambda)$. 

The collision integrals involve also excitations above thermodynamic states. The relevant excitations are neutral particle-hole excitations parametrized by pairs $(p, h)$. Their momentum and energy is $k(p) - k(h)$ and $\omega(p) - \omega(h)$ respectively where
\begin{align}
    k(\lambda) = \lambda - \int {\rm d}\mu\, n(\mu) F(\mu|\lambda), \nonumber \\
    \omega(\lambda) = \lambda^2 - 2 \int {\rm d}\mu\, n(\mu) \mu F(\mu|\lambda). \label{ener-mom}
\end{align}
Here $F(\lambda|\mu)$ is the back-flow function given by
\begin{equation}
    F(\lambda | \mu) = \frac{\theta(\lambda - \mu)}{2\pi} + \int {\rm d}\lambda'\, T(\lambda, \lambda') n(\lambda') F(\lambda' | \mu), \label{SM_backflow}
\end{equation}
where $\theta(\lambda) = 2 \arctan(\lambda/c)$.

In the following we use the notation $\bfp_m = \{p_j\}_{j=1}^m$ and analogously $\bfh_m$ for sets of particles and holes. We drop the subscript $m$ when the size of the set is obvious or immaterial. The dressed momentum and energy are additive in the sense that
\begin{equation}
    k(\bfp, \bfh) = \sum_{j=1}^m \left( k(p_i) - k(h_i)\right),\quad  \omega(\bfp, \bfh) = \sum_{j=1}^m \left( \omega(p_i) - \omega(h_i)\right).
\end{equation}

The filling function $n(\lambda)$ for states following the generalized Gibbs ensemble is specified by the pseudoenergy $\epsilon(\lambda)$. The pseudoenergy follows from the generalized Thermodynamic Bethe Ansatz (gTBA) equation, see Refs.~\onlinecite{2012_Mossel_JPA_45,2012PhRvL.109q5301C}
\begin{equation}
    \epsilon(\lambda) = \epsilon_0(\lambda) - \int {\rm d}\mu T(\lambda, \mu) \log \left(1 + e^{-\epsilon(\mu)} \right).
\end{equation}
where $n(\lambda) = 1/(1 + \exp(\epsilon(\lambda)))$ and $\epsilon_0(\lambda)$ is the bare pseudoenergy. In terms of the ultra-local conserved charges the bare pseudoenergy expands as
\begin{equation}
    \epsilon_0(\lambda) = \sum_k \frac{\beta_k}{k!} q^{(k)}(\lambda), \quad q^{(k)}(\lambda) = \lambda^k,
\end{equation}
with $\beta_k$ the corresponding generalized chemical potentials. At the thermal equilibrium $\beta_0 = - h/T$ and $\beta_2 = 2/T$ and all the other chemical potentials are $0$. In this case $\epsilon(\lambda) = \omega(\lambda)/T$.

\section{Derivation of Collision Integral, $Q^{(n,m)}$}

Let $\rho_{p1}(\lambda)$ be the distribution of occupied rapidities in the first tube.  The scattering induced by the intertube interaction leads to this quantity being time dependent with a form given by
\begin{equation}
    \partial_t \rho_{p1}(\lambda) = \sum_{n,m} Q^{(n,m)}(\lambda).
\end{equation}
It is the aim of this section to derive the form of $Q^{(n,m)}(\lambda)$.

\subsection{Dynamic Structure Function, $S(\omega,k)$}\label{DSF}
We will want to express $Q^{n,m}$ in terms of a form-factor expansion of $S(\omega,k)$, the dynamic structure function (DSF).  We thus begin this section by reviewing this expansion.

The dynamic structure function expands in the number of particle-hole (ph) excitations involved
\begin{equation}
	S(k, \omega) = \sum_{m=1}^{\infty} S^{m}(k, \omega),
\end{equation}
where [\onlinecite{SciPostPhys.1.2.015}]
\begin{equation} \label{S_m_app}
	S^{m}(k,\omega) =  \frac{(2\pi)^2}{(m!)^2} \fint {\rm d}\bfp_m {\rm d} \bfh_m |
	F^{\rho}(\bfp, \bfh)|^2 \delta( k - k(\bfp,\bfh)) \delta(\omega - \omega(\bfp, \bfh)),
\end{equation}
and where $k({\bf p},{\bf h})$ and $\omega({\bf p},{\bf h})$ are defined in Eq.~\ref{ener-mom}. 
In this expression $F^{\rho}(\bfp, \bfh)$ is the thermodynamic form-factor of the density operator $\hat{\rho}(0)$ between state characterized by the particle distribution $\rho_{\rm p}(\lambda)$ and excited state containing particle-hole excitations $(\bfp, \bfh)$ above it. The $2{\rm ph}$ and higher form-factors contain kinematic singularities which require a prescription for the integration over them. This takes form of the Hadamard integral denoted by $\fint$. We discuss the thermodynamic form-factors and details of the regularization in~\ref{SM_ff}. Finally, the integration measure contains the density factors, namely
\begin{equation}
    {\rm d}\bfp_m {\rm d} \bfh_m = \prod_{j=1}^m {\rm d}p_j\, {\rm d}h_j\, \rho_h(p_j)\rho_{\rm p}(h_j).
\end{equation}
We consider now the first two contributions to $S(k,\omega)$ up to the first two orders in $k$.

\subsubsection{1ph contribution} 

The $1{\rm ph}$ contribution to the dynamic structure function, according to~\eqref{S_m_app}, is
\begin{equation}
	S^{1}(k,\omega) =  (2\pi)^2 \int {\rm d}p  {\rm d} h \rho_p(h) \rho_h(p) |F^{\rho}(p, h)|^2 \delta( k - k(p,h)) \delta(\omega - \omega(p, h)).
\end{equation}
In this case the integrals can be completely resolved with the help of the $\delta$-functions. The result is
\begin{equation} \label{S_1_app_full}
	S^{1}(k,\omega) =  (2\pi)^2 \frac{\rho_p(h) \rho_h(p) |F^{\rho}(p, h)|^2}{k'(p) k'(h)|v(p) - v(h)|},
\end{equation}
where $p$ and $h$ are such that $k = k(p) - k(h)$ and $\omega = \omega(p) - \omega(h)$ and $v(\lambda) \equiv \omega'(\lambda)/k'(\lambda)$ is the effective velocity.
We also note that, given the particle-hole-symmetry of the form-factor, $F^{\rho}(p, h) = F^{\rho}(h, p)$, the following relation holds
\begin{equation}
	S^{1}(-k, -\omega) = \frac{\rho_p(p) \rho_h(h)}{\rho_p(h) \rho_h(p)} S^{1}(k, \omega).
\end{equation}
In the thermal equilibrium the ratio of the distribution function is $\exp(-\omega/T)$ and we recover the detailed balance relation.

The expression~\eqref{S_1_app_full} can be further approximated in the small momentum limit. To this end it is convenient to introduce the "central-of-mass" rapidities,
\begin{equation}
	\lambda = \frac{p+h}{2}, \qquad \alpha = p-h.
\end{equation}
such that $v(\lambda) = \omega/k$ and $\alpha = k/k'(\lambda)$. We then find
\begin{equation}
	S^{1}(k,\omega) =  (2\pi)^3 \frac{\rho_p(\lambda- \alpha/2) \rho_h(\lambda + \alpha/2) \rho_t(\lambda)}{v'(\lambda) |k|} (1 + \mathcal{O}(k^2). \label{SM_DSF_1ph}
\end{equation}
 In writing this expression we used that $k'(\lambda) = 2\pi \rho_{\rm tot}(\lambda)$ and that the $1{\rm ph}$ form-factor, in the limit of small excitation is $2\pi \rho_{\rm tot}(\lambda)$, see Eq.~\ref{SM_ff1}.

\subsubsection{2ph contribution}

The $2{\rm ph}$ contribution to the dynamic structure function is 
\begin{equation}
	S^{2}(k,\omega) =  \frac{(2\pi)^2}{4} \fint {\rm d}\bfp_2 {\rm d} \bfh_2 |F^{\rho}(\bfp_2, \bfh_2)\rangle|^2 \delta( k - k(\bfp_2,\bfh_2)) \delta(\omega - \omega(\bfp_2, \bfh_2)).
\end{equation}
We introduce again the change of variables to the "center-of-mass" rapidities, 
\begin{equation}
	\lambda_i = \frac{p_i + h_i}{2} \qquad \alpha_i = p_i - h_i.
\end{equation}
The Jacobian of this transformation equals $1$. Parameters $\alpha_i$ are then fixed by the $\delta$-functions to the following values
\begin{align}
	\alpha_1 = \frac{1}{k'(\lambda_1)}\frac{\omega - v(\lambda_2) k}{v(\lambda_1) - v(\lambda_2)}, \qquad \alpha_2 = \frac{1}{k'(\lambda_2)}\frac{\omega - v(\lambda_1) k}{v(\lambda_2) - v(\lambda_1)}.
\end{align}
Integrating over $\alpha_i$, the dynamic structure function becomes
\begin{equation}
	S^{2}(k,\omega) =  \frac{(2\pi)^2}{4} \fint {\rm d}\bfl_2 \prod_{i=1,2}\frac{\rho_p(\lambda_i - \alpha_i/2) \rho_h(\lambda_i + \alpha_i/2)}{k'(\lambda_i)} \frac{|F^{\rho}(\bfp_2, \bfh_2)|^2}{|v(\lambda_1) - v(\lambda_2)|},
\end{equation}
From the particle-hole symmetry of the form-factor it follows that
\begin{equation}
	S^{2}(-k,-\omega) =  \frac{(2\pi)^2}{4} \fint {\rm d}\bfl_2 \prod_{i=1,2}\frac{\rho_h(\lambda_i - \alpha_i/2) \rho_p(\lambda_i + \alpha_i/2)}{k'(\lambda_i)} \frac{|F^{\rho}(\bfp_2, \bfh_2)|^2}{|v(\lambda_1) - v(\lambda_2)|},
\end{equation}
and the two quantities, unlike the $1{\rm ph}$ case, are directly related to each other only in thermal equilibrium when the detailed balance relation holds, $S^{2{\rm ph}}(-k,-\omega) = e^{-\omega/T} S^{2{\rm ph}}(k,\omega)$.

We use now the expression for the form-factor, see Eq.~\eqref{SM_ff2_DSF}, to find
\begin{equation}
	S^{2}(k,\omega) =  \frac{(2\pi)^2}{2} \fint {\rm d}\bfl_2 \prod_{i=1,2}\frac{\rho_h(\lambda_i - \alpha_i/2) \rho_p(\lambda_i + \alpha_i/2)}{\rho_t(\lambda_i)} \frac{(T^{\rm dr}(\lambda_1, \lambda_2))^2 |v(\lambda_1) - v(\lambda_2)|^3}{(v^* - v(\lambda_1))^2(v^* - v(\lambda_2))^2}, \label{2phDSF}
\end{equation}
where $v^* = \omega/k$. We observe that this expression is in agreement with the detailed balance to all orders in $\omega$.

The formula for the dynamic structure factor can be further expanded in small momenta. The result is of the following structure
\begin{equation}
	S^{2}(k,\omega) = S_0^{2}(\omega/k) + k S_1^{2}(\omega/k) + \mathcal{O}(k^2).
\end{equation}
where
\begin{align}
	S_0^{2}(v^*) &= \frac{(2\pi)^2}{2} \fint {\rm d}\bfl_2 \frac{\rho_p(\bfl) \rho_h(\bfl)}{\rho_t(\bfl)} \frac{(T^{\rm dr}(\lambda_1, \lambda_2))^2 |v(\lambda_1) - v(\lambda_2)|^3}{(v^* - v(\lambda_1))^2(v^* - v(\lambda_2))^2}, \nonumber \\
	S_1^{2}(v^*) &=  \frac{(2\pi)^2}{2} \fint {\rm d}\bfl_2 {\rm sgn}(\lambda_1 - \lambda_2) \frac{\rho_p(\bfl) \rho_h(\bfl)}{\rho_t(\bfl)} \frac{\epsilon'(\lambda_1)}{k'(\lambda_1)} \frac{(T^{\rm dr}(\lambda_1, \lambda_2))^2 (v(\lambda_1) - v(\lambda_2))^2}{(v^* - v(\lambda_1))^2(v^* - v(\lambda_2))}.
\end{align}
The detailed balance implies that at thermal equilibrium the two functions are related through
\begin{equation}
	S_1^{2}(v^*) = \frac{v^*}{2T} S_0^{2}(v^*).
\end{equation}

\subsection{Derivation of $Q^{(1,1)}$}
We define here $\rho_{p1}(\lambda)$ as the occupied rapidity distribution for tube 1.  The indices $1$ and $2$ here always refer to the two tubes.  We first provide a description of how $\rho_{p1}(\lambda)$ changes because of scattering involving 1-particle-hole processes.  Such processes give two additive contributions to $\partial_t \rho_{p1}(\lambda)$: one coming from direct scattering involving the rapidity $\lambda$, ${\partial_t \rho_{p1}(\lambda)}_{\rm direct}$, and one coming from a backflow term, $(\partial_t \rho_{p1}(\lambda))_{\rm backflow}$, which encodes how scattering at other rapidities influences the occupation at $\lambda$ through interactions.

\subsubsection{Direct Contribution}

The direct scattering contribution takes the form
\begin{eqnarray}
{\partial_t \rho_{p1}(\lambda)}_{\rm direct} &=& -L\rho_{p1}(\lambda) \int dh_2 \rho_{p2}(h_2)P^{p_1,p_2}_{\lambda,h_2}.
\end{eqnarray}
Here $P^{p_1,p_2}_{\lambda,h_2}$ is the probability per unit time that a pair of particles $(\lambda,h_2)$ with $\lambda$ in tube 1 and $h_2$ in tube 2 gets scattered into the pair $(p_1,p_2)$ (or alternatively, the probability that a particle-hole pair $(\lambda,p_1)$ in tube 1 and a particle-hole pair $(h_2,p_2)$ in tube 2 are created).  For 1-particle-hole processes, in the case of identical tubes, the kinematics dictates that $(p_1,p_2)=(h_2,\lambda)$, i.e. the scattering between the two tubes exchanges rapidities.

$P^{p_1,p_2}_{\lambda,h_2}$ is given by 
\begin{eqnarray}\label{1ph}
P^{p_1,p_2}_{\lambda,h_2} &=& \bigg((R^{out})^{p_1,p_2}_{\lambda,h_2} - (R^{in})^{\lambda,h_2}_{p_1,p_2}\frac{\rho_{h1}(\lambda)\rho_{h2}(h_2)}{\rho_{p1}(p_1)\rho_{p2}(p_2)} \bigg),
\end{eqnarray}
where $(R^{out})^{p_1,p_2}_{\lambda,h_2}$ is the rate at which two particles $(\lambda,h_2)$
scatter into $(p_1,p_2)$ while $(R^{in})^{\lambda,h_2}_{p_1,p_2}$ marks the reverse process of $(p_1,p_2)$ scattering into $(\lambda,h_2)$.  This last rate is weighted by the particle, $\rho_p$, and hole, $\rho_h$, densities: the ability to scatter into $(\lambda,h_2)$ is determined by the hole distributions, $\rho_{h1}(\lambda),\rho_{h2}(h_2)$, while the ability to scatter out of $(p_1,p_2)$ is determined by $\rho_{p1}(p_1),\rho_{p2}(p_2)$.

$(R^{out})^{p_1,p_2}_{\lambda,h_2}$ is given from Fermi's golden rule as 
\begin{eqnarray}
(R^{out})^{p_1,p_2}_{\lambda,h_2} &=& \frac{8\pi^2 m V_0^2}{\hbar^3 L}\int dp_1dp_2 A^2(\Delta k)\delta(\omega_1(p_1)+\omega_2(p_2)-\omega_1(\lambda)-\omega_2(h_2))\cr\cr
&&\hskip .5in \times \delta(k_1(p_1)+k_2(p_2)-k_1(\lambda)-k_2(h_2)) \rho_{h1}(p_1)\rho_{h2}(p_2) \cr\cr
&&\hskip .5in \times |F^{\rho_{1}}(p_1;\lambda) F^{\rho_{2}}(p_2;h_2)|^2 \cr\cr
&=& \frac{4\pi m V_0^2A^2(\Delta k)}{\hbar^3 L}\frac{\rho_{h1}(p_1)\rho_{h2}(p_2)}{|\rho_{T2}(p_2)\omega_1'(p_1)-\rho_{T1}(p_1)\omega_2'(p_2)|} \cr\cr
&& \hskip 1in \times |F^{\rho_{1}}(p_1;\lambda) F^{\rho_{2}}(p_2;h_2)|^2\bigg|_{p_1,p_2~{\rm fixed~by~ener.-mom.~ cons.}}.\cr\cr&&
\end{eqnarray}
Here $p_1$ and $p_2$ are fixed by energy-momentum conservation, i.e.,
\begin{eqnarray}
0 = k_1(p_1) - k_1(\lambda) + k_2(p_2) - k_2(h_2);\cr\cr
0 = \omega_1(p_1) - \omega_1(\lambda) + \omega_2(p_2) - \omega_2(h_2),
\end{eqnarray}
where $\omega_i(\lambda)$ and $k_i(\lambda)$ are the dressed energy and momentum (see Section S1).
In the above we have used the relation between the momentum of an excitation at rapidity $\lambda$ and the total density of states at this same rapidity:
\begin{eqnarray}
 \rho_{Ti}(\lambda) \equiv \rho_{pi}(\lambda)+\rho_{hi}(\lambda) = \frac{k_i'(\lambda)}{2\pi}.
\end{eqnarray}
In the case of the two tubes being identical, energy-momentum conservation amounts to 
\begin{equation}
    p_1 = h_2; ~~~ p_2 = \lambda.
\end{equation}
The quantity
\begin{equation}
F^{\rho_{1}} (p_1;\lambda) \equiv L\langle p_1,\lambda|\hat\rho_{1}(x=0)|\rho_{p1}(\lambda)\rangle,
\end{equation}
gives the 1-ph (one particle-hole) dimensionless density form factor on tube 1 involving one particle at $p_1$ and one hole at $\lambda$.
This expression for $(R^{out})^{p_1,p_2}_{\lambda,h_2}$ has the dimensionful factors appearing as a prefactor whereas the rapidities, $\lambda,h_2,p_1,p_2$, and thermodynamic densities $\rho_{p/h i}$ and dressed energies/momentum $\omega_i,k_i$ are in rationalized units ($\hbar=1,m=1/2$).

Similarly $(R^{in})^{\lambda,h_2}_{p_1,p_2}$ is given by Fermi's golden rule to be
\begin{eqnarray}
(R^{in})^{\lambda,h_2}_{p_1,p_2} &=& \frac{8\pi^2 V_0^2}{L\hbar^3}\int dp_1dp_2 A^2(\Delta k)\delta(\omega_1(p_1)+\omega_2(p_2)-\omega_1(\lambda)-\omega_2(h_2))\cr\cr
&&\hskip .5in \times \delta(k_1(p_1)+k_2(p_2)-k_1(\lambda)-k_2(h_2)) \rho_{p1}(p_1)\rho_{p2}(p_2) \cr\cr
&&\hskip .5in \times | F^{\rho_{1}}(\lambda;p_1) F^{\rho_{2}}(h_2;p_2)|^2 \cr\cr
&=& \frac{4\pi m V_0^2 A^2(\Delta k)}{L\hbar^3}\frac{\rho_{p1}(p_1)\rho_{p2}(p_2)}{|\rho_{T2}(p_2)\omega_1'(p_1)-\rho_{T1}(p_1)\omega_2'(p_2)|} \cr\cr&&
\hskip 1in \times |F^{\rho_{1}}(p_1;\lambda) F^{\rho_{2}}(p_2;h_2)|^2\bigg|_{p_1,p_2~{\rm fixed~by~ener.-mom.~ cons.}}.\cr\cr&&
\end{eqnarray}
Putting everything together, the expression for ${\partial_t \rho_{p1}(\lambda)}_{\rm direct}$ is
\begin{eqnarray}
{\partial_t \rho_{p1}(\lambda)}_{\rm direct}  &=& \frac{4\pi V_0^2  m}{\hbar^3}\int dh_2 A^2(\Delta k)
\frac{| F^{\rho_{1}}(p_1;\lambda) F^{\rho_2}(p_2;h_2)|^2}{|\rho_{T2}(p_2)\omega'_1(p_1)-\rho_{T1}(p_1)\omega'_2(p_2)|}\cr\cr
&& \hskip -1.5 in \times (\rho_{h1}(\lambda)\rho_{h2}(h_2)\rho_{p1}(p_1)\rho_{p2}(p_2)-\rho_{h1}(p_1)\rho_{h2}(p_2)\rho_{p1}(\lambda)\rho_{p2}(h_2))\bigg|_{p_1,p_2~{\rm fixed~by~ener.-mom.~ cons.}}.\cr &&
\end{eqnarray}
We can see that this vanishes if the hole and particle densities $\rho_{pi}$ and $\rho_{hi}$ are the same for both tubes.

We want to express ${\partial_t \rho_{p1}(\lambda)}_{\rm direct}$ in terms of the DSF of the second tube.  The 1-ph contribution to this DSF, $S^{1}_2(k,\omega)$ is given by (see \ref{DSF})
\begin{eqnarray}
S^{1}_2(k,\omega) &\equiv& S^{1}(k_2(p_2)-k_2(h_2),\omega_2(p_2)-\omega_2(h_2))=S^{1}(p_2,h_2)\cr\cr
&=&2\pi \frac{|F^{\rho_2}(p_2;h_2)|^2\rho_{p2}(h_2)\rho_{h2}(p_2)}{|\rho_{T2}(h_2)\omega'_2(p_2) - \rho_{T2}(p_2)\omega'_2(h_2)|}.
\end{eqnarray}
This then allows us to write
\begin{eqnarray}
(\partial_t \rho_{p1}(\lambda))_{\rm direct} &=& \frac{16 E_F}{\hbar \pi^2} \gamma_{inter}^2 \int dh_1 dp_1 A^2(\Delta k)\delta(\lambda-h_1)|F^{\rho_1}(p_1;h_1)|^2\cr\cr
&& \hskip -1in \times (\rho_{h1}(h_1)\rho_{p_1}(p_1)S^{1}_2(h_2;p_2)
-\rho_{h1}(p_1)\rho_{p_1}(h_1)S^{1}_2(p_2;h_2))|_{p_2,h_2~{\rm fixed~by~ener.-mom.~ cons.}};\cr\cr
\gamma_{inter} &\equiv& \frac{V_0 m}{n\hbar^2}; ~~~
E_F \equiv \frac{\hbar^2 k_F^2}{2 m}, ~~k_F \equiv \frac{N}{2L},
\end{eqnarray}
where we now have expressed the dimensionful prefactor in terms of $\gamma_{inter}$ which marks the dimensionless strength of the couplings between tube and the Fermi energy, $E_F$, of the gas.  $E_F/\hbar$ then sets the timescale for the evolution of the gas.  We have also done a change of variables in the integration to those of tube 1's hole ($h_1$) and particle ($p_1$).  If the hole and particle dummy variables are interchanged (i.e. $p_i \leftrightarrow h_i$) and the momentum $k$ and energy $\omega$ at which $S_2$ is evaluated is taken to be that of tube 1, the r.h.s. of this expression becomes the $Q_0^{(1,1)}$ of Eq. 4 of the main text.

\subsubsection{Backflow Contribution}

The backflow contribution takes the form
\begin{eqnarray}
  (\partial_t \rho_{p1}(\lambda))_{\rm backflow} &=& \sum_{s_1's_2'} (R^{out})^{p_1,p_2}_{h_1h_2}\delta \rho_{p1}(\lambda|\{p_1\};\{h_1\}),
\end{eqnarray}
where $\delta \rho_{p1}$ is the change in $\rho_{p1}$ due to scattering events involving the creation of a set of particles $\{p_1\}$ and holes $\{h_1\}$ in tube 1:\footnote{This form of $\delta \rho_{p1}$ can be read off from Eq. 8.7 of Chapter 1 in  \onlinecite{KBI}}
\begin{eqnarray}
\delta \rho_{p1}(\lambda|\{p_1\};\{h_1\})  &=& \frac{1}{L} \partial_{\lambda}(F(\lambda| \{p_1\};\{h_1\})n_1(\lambda)).
\end{eqnarray}
Here the shift function $F(\lambda|\{p\}; \{h\})$ satisfies
\begin{equation}
    F(\lambda| \{p_1\};\{h_1\}) = \sum_{p_1\in\{p_1\}}F(\lambda|p_1) -\sum_{h_1\in\{h_1\}}F(\lambda|h_1),
\end{equation}
with $F(\lambda|\mu)$ defined in~\eqref{SM_backflow}.
Unlike the direct contribution to scattering, the backflow only depends on $R^{out}$: one is summing over all processes (as indexed in the first term, say, by the holes $(h_1,h_2)$ that influence through interactions the density at $\lambda$).  This indexing is complete and to include $R^{in}$ would be to double-count.  

To provide a compact expression for $(\partial_t \rho_{p1}(\lambda))_{\rm backflow}$, we note that the shift function depends in a linear fashion on its p-h arguments. 
Using this we can write the 1-ph contribution to $(\partial_t \rho_{p1}(\lambda))_{\rm backflow}$ as
\begin{eqnarray}
(\partial_t \rho_{p1}(\lambda))_{\rm backflow} &=& \frac{4\pi m V_0^2}{\hbar^3} \int dh_1dh_2 A^2(\Delta k)\frac{| F^{\rho_{1}}(p_1;h_1) F^{\rho_2}(p_2;h_2)|^2 }{|\rho_{T2}(p_2)\omega'_1(p_1)-\rho_{T1}(p_1)\omega'_2(p_2)|}\cr\cr
&& \hskip -1.0in \times \rho_{h1}(p_1)\rho_{h2}(p_2)\rho_{p1}(h_1)\rho_{p2}(h_2) 
 \partial_{\lambda}(F(\lambda|\{p_1\};\{h_1\})n_1(\lambda))\bigg|_{p_1,p_2~{\rm fixed~by~ener.-mom.~ cons.}}    \cr\cr
&=& \frac{4\pi m V_0^2}{\hbar^3} \int dh_1dh_2 A^2(\Delta k)\frac{| F^{\rho_{1}}(p_1;h_1) F^{\rho_2}(p_2;h_2)|^2 }{|\rho_{T2}(p_2)\omega'_1(p_1)-\rho_{T1}(p_1)\omega'_2(p_2)|}\cr\cr
&& \hskip -.6in \times (\rho_{h1}(h_1)\rho_{h2}(h_2)\rho_{p1}(p_1)\rho_{p2}(p_2)-\rho_{h1}(p_1)\rho_{h2}(p_2)\rho_{p1}(h_1)\rho_{p2}(h_2))\cr\cr
&& \hskip -.6in \times  \partial_{\lambda}(F(\lambda|h_1;\{\})n_1(\lambda))\bigg|_{p_1,p_2~{\rm fixed~by~ener.-mom.~cons.}},
\end{eqnarray}
where in the last line we have assumed the form factors are p-h symmetric and have swapped dummy particle indices with dummy hole indices.  We again see that if the tubes are identical, the contribution to $\partial_t \rho_{p1}(\lambda)$ goes to zero.

Like with the direct contribution, we will now write the backflow contribution in terms of the 1-particle-hole contribution to the dynamic structure factor. 
This then allows us to write
\begin{eqnarray}
(\partial_t \rho_{p1}(\lambda))_{\rm backflow} &=& \frac{16E_F}{\hbar \pi^2} \gamma_{inter}^2 \int dh_1 dp_1 A^2(\Delta k)\partial_{\lambda}(F(\lambda|h_1;\{\})n(\lambda))
|F^{\rho_1}(p_1;h_1)|^2\cr\cr
&& \hskip -1.0in \times (\rho_{h1}(h_1)\rho_{p1}(p_1)S^{1ph}_2(h_2;p_2) - \rho_{h1}(p_1)\rho_{p1}(h_1)S^{1ph}_2(p_2;h_2))|_{p_2,h_2~{\rm fixed~by~ener.-mom.~ cons.}}. \cr&&
\end{eqnarray}
We can then write both the direct and backflow contributions to $\partial_t \rho_{p1}(\lambda)$ as
\begin{eqnarray}\label{full_eq}
\partial_t \rho_{p1}(\lambda) &=& \frac{16E_F}{\hbar \pi^2} \gamma_{inter}^2 \int dh_1 dp_1 A^2(\Delta k) R(\lambda,h_1)
|F^{\rho_1}(p_1;h_1)|^2\cr\cr
&& \hskip -.5in \times (\rho_{h1}(h_1)\rho_{p1}(p_1)S^{1}_2(h_2;p_2) - \rho_{h1}(p_1)\rho_{p1}(h_1)S^{1}_2(p_2;h_2))|_{p_2,h_2~{\rm fixed~by~ener.-mom.~ cons.}}, \cr
&&
\end{eqnarray}
where the kernel $R(\lambda,h_1)$ is given by
\begin{equation}\label{backflow_kernel}
    R(\lambda,h_1) = \delta(\lambda-h_1) + \partial_{\lambda}(F(\lambda|h_1;\{\})n(\lambda)).
\end{equation}

\subsubsection{Vanishing of $Q^{(1,1)}$}

As we have already noted, when the two tubes have identical distributions, $Q^{(1,1)}$ vanishes.  When the tubes are initialized where they are inequivalent, $Q^{(1,1)}$ will lead to a process where the distributions approach one another. In the case when the particle densities in both the tubes are the same (as considered in the main text) the stationary state is simply
$n_1(\lambda)=n_2(\lambda)$,
and thus the dispersion relations for both tubes are the same:
\begin{equation}
    \omega_1(\lambda) = \omega_2(\lambda); ~~ k_1(\lambda) = k_2(\lambda).
\end{equation}
Consequently, energy-momentum conservation in Eq.~\ref{full_eq} gives $p_2=h_1$ and $p_1=h_2$.
This condition combined with the fact that now $\rho_{p1}(\lambda) = \rho_{p2}(\lambda)$ and $\rho_{h1}(\lambda) = \rho_{h2}(\lambda)$ leads to $\partial_t\rho_{p1}(\lambda)=0$, i.e.  a stationary point.  We emphasize that in general the fixed point arrived at by $Q^{(1,1)}$ processes will not correspond to a thermal distribution. 

\subsection{Derivation of $Q^{(n,m)}$}

The derivation of expressions for $Q^{(n, m)}$ is analogous to that of $Q^{(1, 1)}$. The difference is that now we consider processes involving $n$ particle-hole pairs in the first tube and $m$ particle-hole pairs in the second tube. Summing over the phase space of the second tube again leads to the dynamic structure function, this time the contribution to the DSF involving $m$ particle-hole pairs. On the other hand, in the first tube we have to track all the excitations that modify the specified rapidity $\lambda$. For $m_1$ pairs it leads to a term proportional to $\sum_{j=1}^{n} \delta( \lambda - p_i)$ for processes in which there is particle created at $\lambda$ and a term proportional to $\sum_{j=1}^{n} \delta( \lambda - h_i)$ for processes in which there is a hole being created at $\lambda$. Finally, when writing the phase space for $n$ particle-hole excitations we overcount by allowing the excitations that differ only in the labelling of particles and labelling of holes. We take this into account be dividing the rate by $(n!)^2$. The rate then becomes
\begin{align}
    Q^{(n,m)}_0(\lambda) &= \frac{1}{(n!)^2} \sum_{i=1}^{n} \int \prod_{i = 1}^{n} d p_i d h_i \delta(\lambda - h_i) A^2(k_1)  \left|f(\{ p_i, h_i \})\right|^2 \nonumber \\
    &\times \left[ \rho_p(p_i) \rho_h(h_i) S^{m}(- k_1,- \omega_1) - \rho_h(p_i) \rho_p(h_i) S^{m}( k_1,  \omega_1) \right].
\end{align}
Here we have made explicit the effects of the dipole-dipole interaction between tubes as encoded in the Fourier transform, $A(k)$.
We have used the particle-hole symmetry of the form factors $F^{\rho_1}$ to write the contribution to $Q^{(n,m)}$ in terms of the holes being created at $\lambda$.  In this way, we obtain Eq.~4 of the main text recognizing that each term of the sum $\sum_{i=1}^{n}$ contributes equally.  We can thus replace the sum by simply multiplying the rate by $n$. 

The quantity $Q^{(n,m)}_0(\lambda)$ accounts for the direct scattering. The back-flow contributions is then included in the analogous way as for $(1,1)$ processes by the integral operator $R$ given in Eq.~\ref{backflow_kernel}  as stated in the main text.

\section{Entropy production}

We discuss now the process of thermalization from the point of view of entropy production. The entropy initially increases as the result of the Bragg splitting. Following that, the entropy increases further as the non-equilibrium Bragg distribution approaches the thermal one. To compute the non-equilibrium entropy we adopt the diagonal entropy~[\onlinecite{2011AnPhy.326..486P}] to the Boltzmann picture of the time evolution. The diagonal entropy is defined solely from the diagonal elements, in the instantaneous eigenenergy basis, of the density matrix, $S_d = \sum_n \rho_{nn} \log \rho_{nn}$ and captures the entropy production in time-dependent processes in Hamiltonian systems. If the system is in thermal equilibrium then $S_d = S_N$ where $S_N = {\rm tr}\rho \log \rho$ is the von Neumann entropy. 

\begin{figure}
    \centering
    \includegraphics[scale=0.6]{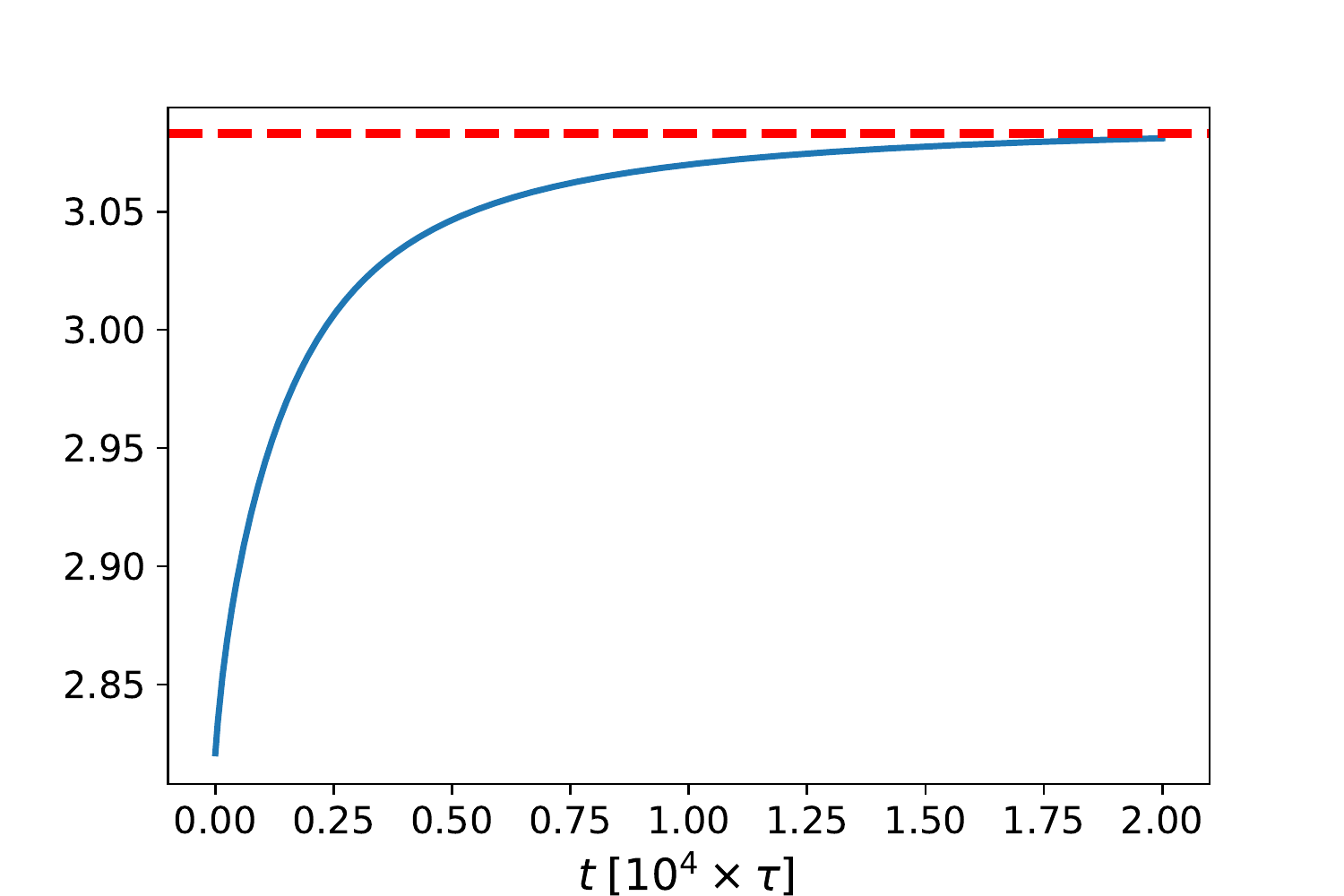}
    \caption{The evolution of the total entropy of the system for the non-equilibrium protocol considered in the main text. The dashed red line is the limiting value acquired at large times.}
    \label{fig:entropy}
\end{figure}

In our setup, the two tubes are coupled perturbatively and therefore the state of the system is the product state of two representative state in the two tubes. The entropy is then a sum of two identical contributions given that the states in both tubes are the same. In the thermodynamic limit, through standard arguments, the entropy of a single tube is then equal to the Yang-Yang entropy and the total entropy at time $t$ is $S(t) = S_1(t) + S_2(t)$ where the time dependence enters through the time dependence of the particles distribution, $S_i(t) = S[\rho_{\rm p, i}(t)]$.
\begin{equation}
    S[\rho_{\rm p}] = \int {\rm d}\lambda \rho_{\rm tot}(\lambda) G(n(\lambda)),
\end{equation}
with $G(n) = n\log n - (1 - n)\log(1 - n)$. Note that even though the system is bosonic, the entropy has a fermionic form. In Fig.~\ref{fig:entropy} we show the time evolution of the entropy for the system considered in the main text.

\section{Linearized Collision Integral}

We derive here a linearized form of the collision integral $Q^{(n,m)}$. We are considering here only the case of equal distributions in both tubes. In such case $Q^{(1,1)}$ is identically equal to zero and the following analysis applies to collision integrals of higher orders. To derive a linearized collision integral it is convenient to use the expression for the dynamic structure function in the spectral representation in formula for $Q^{(n, m)}_0(\lambda)$. We find
\begin{align}
    Q^{(n,m)}_0(\lambda) &= \frac{n}{(n!)^2} \frac{1}{(m!)^2}\int \prod_{i = 1}^{n} d p_{i,1} d h_{i,1} \prod_{i = 1}^{m} d p_{i,2} d h_{i,2}\delta(\lambda - p_{1,1}) A^2(k_1)  \nonumber \\
    &\times \left|F^{\rho_1}(\{ p_{i,1}, h_{i,1} \})\right|^2 \left|F^{\rho_2}(\{ p_{i,2}, h_{i, 2} \})\right|^2 \delta(k_1 - k_2) \delta(\omega_1 - \omega_2)\nonumber \\
    &\times \left[ \rho_p(p_{i,1}) \rho_h(h_{i,1}) \rho_p(p_{i,2}) \rho_h(h_{i,2}) - \rho_h(p_{i,1}) \rho_p(h_{i,1}) \rho_h(p_{i,2}) \rho_p(h_{i,2}) \right].
\end{align}
In the vicinity of the equilibrium we parametrize the gTBA pseudoenergy as $\epsilon(\lambda) = \epsilon_{\rm th}(\lambda) + \delta \epsilon(\lambda)$. We then find
\begin{equation}
    \frac{\rho_p(p_{i,1}) \rho_h(h_{i,1}) \rho_p(p_{i,2}) \rho_h(h_{i,2})}{\rho_h(p_{i,1}) \rho_p(h_{i,1}) \rho_h(p_{i,2}) \rho_p(h_{i,2})} = \exp\left(\sum_{i=1}^{m_1}(\delta\epsilon(p_{i,1}) - \delta\epsilon(h_{i,1})) + \sum_{i=1}^{m_2}(\delta\epsilon(p_{i,2}) - \delta\epsilon(h_{i,2}))\right),
\end{equation}
which, for small $\delta\epsilon(\lambda)$ can be further expanded leading to
\begin{align}
    Q^{(n,m)}_0(\lambda) &= \int \prod_{i = 1}^{n} d p_{i,1} d h_{i,1} \prod_{i = 1}^{m} d p_{i,2} d h_{i,2} P(\{p_{i,1}, h_{i, 1}\},\{p_{i,2}, h_{i, 2}\})  \nonumber \\
    &\times \sum_{i=1}^{n}\delta(\lambda - p_{i, 1})\left(\sum_{i=1}^{n}(\delta\epsilon(p_{i,1}) - \delta\epsilon(h_{i,1})) + \sum_{i=1}^{m}(\delta\epsilon(p_{i,2}) - \delta\epsilon(h_{i,2})) \right),
\end{align}
where we have defined 
\begin{align}
    P(\{p_{i,1}, h_{i, 1}\},\{p_{i,2}, h_{i, 2}\}) &= \frac{1}{(n!)^2} \frac{1}{(m!)^2} A^2(k_1) \rho_p(p_{i,1}) \rho_h(h_{i,1}) \rho_p(p_{i,2}) \rho_h(h_{i,2}) \nonumber \\
    &\times \left|F^{\rho_1}(\{ p_{i,1}, h_{i,1} \})\right|^2 \left|F^{\rho_2}(\{ p_{i,2}, h_{i, 2} \})\right|^2 \delta(k_1 - k_2) \delta(\omega_1 - \omega_2).
\end{align}
In the leading order in $\delta\epsilon(\lambda)$ we can take the distribution functions $\rho_p(\lambda)$ and $\rho_h(\lambda)$ to be thermal. Then the collision integral $Q^{(n,m)}_0(\lambda)$ in this approximation can be viewed as a result of action of a linear operator on function $\delta\epsilon(\lambda)$. To this end, we introduce $\mathbf{Q}_0^{(n, m)}(\lambda, \mu)$ such that
\begin{equation}
    Q^{(n, m)}_0(\lambda) = \int {\rm d}\mu\, \mathbf{Q}_0^{(n, m)}(\lambda, \mu) \delta\epsilon(\mu), 
\end{equation}
where
\begin{align}
    \mathbf{Q}_0^{(n, m)}(\lambda, \mu) & = \int \prod_{i = 1}^{n} d p_{i,1} d h_{i,1} \prod_{i = 1}^{m} d p_{i,2} d h_{i,2}
    P(\{p_{i,1}, h_{i, 1}\},\{p_{i,2}, h_{i, 2}\}) \sum_{i=1}^{n}\delta(\lambda - p_{i, 1}) \nonumber \\
    &\times \left(\sum_{i=1}^{n}\left(\delta(p_{i,1} - \mu) - \delta(h_{i,1} - \mu)\right) + \sum_{i=1}^{m}\left(\delta(p_{i,2} - \mu) - \delta(h_{i,2} - \mu)\right)\right).
\end{align}
The operator $\mathbf{Q}_0^{(n, m)}$ itself is not symmetric, however the combination $\bar{\mathbf{Q}}_0^{(n, m)} = \mathbf{Q}_0^{(n, m)} + \mathbf{Q}_0^{(m, n)}$ appearing in the evolution equation is. It is straightforward to see this directly from the definition of $\mathbf{Q}_0^{(n, m)}$. The first term containing $\delta(\lambda - p_{i,1})\delta(p_{i, 1} - \lambda)$ is clearly symmetric. The second term is symmetric because the expression is invariant under relabelling particles and holes. Therefore the first two terms are symmetric on their own within $\mathbf{Q}_0^{(n, m)}$. Exchanging $\lambda$ and $\mu$ in the third term leads to a corresponding term in $\mathbf{Q}_0^{(m, n)}$. Similarly for the fourth term after an additional relabelling of particles and holes. 

The time evolution in the vicinity of a thermal state depends on the spectrum of  operators $\bar{\mathbf{Q}_0}^{(n,m)}$. The situation is the most transparent for large $c$. The leading order is then controlled by the $2$ particle-hole form-factor which behaves like $1/c^2$ and the effect of various dressings is subleading and can be neglected. We then find the following equation for the deviation $\delta \epsilon(\lambda, t)$ of the gTBA pseudoenergy from the thermal one,
\begin{equation}
    \frac{1}{2\pi} n(\lambda)(1 - n(\lambda))\partial_t \delta \epsilon(\lambda, t) = - \sum_{n, m} \int {\rm d}\mu\, \mathbf{Q}_0^{(n, m)}(\lambda, \mu) \delta \epsilon(\mu, t),
\end{equation}
with the leading processes $(1,2)$ and $(2,1)$ determined by $\bar{\mathbf{Q}}_0^{(1,2)}$.  

The late time dynamics depends on the smallest non-zero eigenvalue. We estimate this eigenvalue by numerically evaluating the matrix elements of $\bar{\mathbf{Q}}_0^{(1, 2)}$ in the basis formed from the first $n$ relevant ultra-local conserved charges. Namely, we form a finite-size matrix with elements 
\begin{equation}
    \left(\bar{\mathbf{Q}}_0^{(1, 2)}\right)_{ij} = \int{\rm d}\mu {\rm d}\lambda\, q_{2(i+2)}(\lambda) \bar{\mathbf{Q}}_0^{(1, 2)}(\lambda, \mu) q_{2(j+1)}(\mu), \quad i,j = 1, 2, \dots, n.
\end{equation}
We then numerically diagonalize this matrix. The resulting spectrum is reported in the main text.

\section{Discussion of Form Factors in Thermodynamic Limit} \label{SM_ff}

The thermodynamic form-factors are matrix elements of an operator between thermodynamic state of the system. Their construction assumes that the state of the system is characterized by distributions of particles $\rho_{\rm p}(\lambda)$ and a number of particle hole excitations. For the case of the Lieb-Liniger model the thermodynamic form-factors of the density operator were developed in~[\onlinecite{Smooth_us,2018JSMTE..03.3102D,10.21468/SciPostPhys.6.4.049,milosz_2021}] and are defined as
\begin{equation}
    F^{\rho}(\bfp, \bfh) = \langle \rho_{\rm p} | \hat{\rho}(0) | \rho_{\rm p}; \bfp, \bfh\rangle.
\end{equation}
We note that this definition assumes normalization of the states. Below we present formulas for $1{\rm ph}$ and $2{\rm ph}$ form-factors in the limit of small excitations ($p_i \sim h_j$) which are relevant for the evaluation of the collision integrals $Q^{(1,2)}$ and $Q^{(2,1}$,
\begin{align}
    F^{\rho}(p, h) &= 2\pi \rho_{\rm tot}(h) + \left(\dots\right), \label{SM_ff1} \\
    F^{\rho}(p_1, p_2, h_1, h_2) &= k T^{\rm dr}(h_1, h_2)\left( \frac{1}{k'(h_1)(p_1 - h_1)} + \frac{1}{k'(h_2)(p_2 - h_2)}\right. \nonumber \\
    &\left. + \frac{1}{k'(h_1)(p_2 - h_1)} + \frac{1}{k'(h_2)(p_1 - h_2)} + \left( \dots\right)\right). \label{SM_ff2}
\end{align}
In the second expression $k = k(p_1) + k(p_2) - k(h_1) - k(h_2)$ is the momentum of the excited state, $T^{\rm dr}(h_1, h_2)$ is the dressed scattering kernel 
\begin{equation}
    T^{\rm dr}(\lambda, \lambda') = T(\lambda, \lambda') + \int {\rm d}\lambda'' n(\lambda'') T(\lambda, \lambda'') T^{\rm dr}(\lambda'', \lambda),
\end{equation}
and $\left(\dots\right)$ stand for subleading corrections in the particle-hole differences $p_i - h_j$.

The single particle-hole pair form-factor is a regular function of the excitations. Instead, the two particle-hole pair form-factor has simple poles when any particle coincides with any hole. For computations of the collision integrals and dynamic structure factor, the form-factors are squared leading to singularities which have to be regularized. The proper regularization is the Hadamard one given by the following formula [\onlinecite{Pozsgay_2010,10.21468/SciPostPhys.6.4.049,CortsCubero2019}]:
\begin{equation}
    \int_a^b {\rm d}x \frac{f(x)}{x^2} = \lim_{\epsilon \rightarrow 0} \left( \int_a^b{\rm d}x \frac{f(x)}{x^2} \theta(|x| - \epsilon) - \frac{2 f(0)}{\epsilon}\right).
\end{equation}

Given the particle-hole symmetry of the form-factors, expressions~\eqref{SM_ff1} and~\eqref{SM_ff2} can be symmetrised by introducing the "center-of-mass" rapidity $\lambda_{ij} = (p_i+h_j)/2$ and a bare momentum $\alpha_{ij} = p_i-h_j$. The particle-hole symmetry of the form-factors implies then that they are even functions of $\alpha_{ij}$. This gives then the following symmetrised formulas
\begin{align}
    F^{\rho}(p, h) &= 2\pi \rho_{\rm tot}(\lambda) + \mathcal{O}(\alpha^2), \\
    F^{\rho}(p_1, p_2, h_1, h_2) &= \frac{k^2 g(p_1, p_2, h_1, h_2)}{(p_1 - h_1)(p_1 - h_2)(p_2 - h_1)(p_2 - h_2)} + \mathcal{O}(\alpha_{ij}^2), 
\end{align}
where
\begin{equation}
    g(p_1, p_2, h_1, h_2) = \frac{T^{\rm dr}(\lambda_{11}, \lambda_{22})}{k'(\lambda_{11}) k'(\lambda_{22})}(p_1 - h_1)(p_2 - h_2) + \frac{T^{\rm dr}(\lambda_{12}, \lambda_{21})}{k'(\lambda_{12}) k'(\lambda_{21})}(p_1 - h_1)(p_2 - h_2).
\end{equation}
Furthermore, in situations where the phase space can be restricted such that $p_1 - h_1$ and $p_2 - h_2$ are both the smallest excitations, the two particle-hole pair form-factor further reduces to
\begin{equation}
    F^{\rho}(p_1, p_2, h_1, h_2) = \frac{k^2 T^{\rm dr}(\lambda_{11}, \lambda_{22}) }{k'(\lambda_{11}) k'(\lambda_{22})(p_1 - h_1)(p_2 - h_2)}. \label{SM_ff2_DSF}
\end{equation}

\bibliography{bib}